\documentclass[11pt]{article}

\usepackage[final]{acl}

\usepackage{times}
\usepackage{latexsym}
\usepackage[T1]{fontenc}
\usepackage[utf8]{inputenc}
\usepackage{microtype}
\usepackage{inconsolata}
\usepackage{graphicx}
\usepackage{booktabs}
\usepackage{multirow}
\usepackage{amsmath}
\usepackage{amssymb}
\usepackage{xspace}
\usepackage{url}
\usepackage{tikz}
\usepackage{textgreek}
\usepackage[absolute,overlay]{textpos}

\usetikzlibrary{positioning,arrows.meta,shapes.geometric,fit,backgrounds}
\usepackage{tabularx}
\usepackage{comment}
\usepackage{booktabs}
\usepackage{multirow}
\newcommand{\skillpool}{34{,}396\xspace}
\newcommand{\hitk}[1]{\textsc{Hit}@#1\xspace}
\newcommand{\recallk}[1]{\textsc{Recall}@#1\xspace}

\newcommand{\mrr}{\textsc{MRR}\xspace}

\newcommand{\vs}{vs.\xspace}

\title{When Synthetic Data Hurts: On Catastrophic Forgetting in Skill Retrieval for LLM Agents}

\author{
  \textbf{Syed Shariyar Murtaza}\textsuperscript{1} \quad
  \textbf{Yifan Nie}\textsuperscript{1} \quad
  \textbf{Utkarsh Soni}\textsuperscript{1} \\
  \textbf{Eugene Wen}\textsuperscript{1} \quad
  \textbf{Arvid Frydenlund}\textsuperscript{1} \\
  \textsuperscript{1}Manulife, 200 Bloor St E, Toronto, ON M4W 1E5, Canada \\
  \texttt{\{syed\_shariyar\_murtaza, yifan\_nie, utkarsh\_soni,} \\
  \texttt{eugene\_wen, arvid\_frydenlund\}@manulife.com}
}
\usepackage{newunicodechar}
\newunicodechar{≥}{\geq}

\usepackage{siunitx}

\begin{document}
\maketitle
\begin{textblock*}{18cm}(1.5cm,28.5cm) 
\centering
\footnotesize
   Proceedings of the 2026 Conference on Empirical Methods in Natural Language Processing: Industry Track, 
   \\ Oct 26-29, 2026, ©2026 Association for Computational Linguistics
\end{textblock*}

\begin{abstract}
LLM agents increasingly rely on external skills retrieved at runtime, making skill selection from large repositories a critical challenge. We present a production skill router over 34,396 skills and a large-scale study of skill retrieval using limited real supervision and synthetic data. We found that the synthetic-data fine-tuning improves in-distribution retrieval but it causes catastrophic forgetting on real and out-of-distribution (OOD) data. We evaluate several forgetting mitigation fine-tuning approaches inspired by continual learning, including embedding-anchor regularization, Learning without Forgetting (LwF), Elastic Weight Consolidation (EWC), and L2-initialization. The results show that these approaches not only retain the performance on OOD skills retrieval but also improve the retrieval on synthetic in-distribution skills by 13.98\% for 0.6B Qwen retriever and reranker. Our results provide a practical benchmark and a robust fine-tuning recipe for scarce, multi-positive supervision.

\end{abstract}

\section{Introduction}
\label{sec:intro}

Modern LLM-driven agents increasingly rely on \emph{skills}: short, agent-readable procedures retrieved at run time and injected into the agent's context to solve specific tasks~\citep{PLACEHOLDER_anthropic_skills,PLACEHOLDER_react}. These tasks span scientific (detecting exoplanets and computing orbital periods), engineering (HVAC control), financial (SEC report generation), software (debugging a Node.js app) and many other domains, and they require specific skills for end-to-end completion; e.g., a fraud-detection task may depend on fuzzy matching, database search, PDF reading, and Excel analysis.  As skill registries grow, retrieving the right skills for a task becomes hard. The common information retrieval practice is to use sparse retrieval algorithms, such as BM25 and dense vector retrieval models to match on lexical and semantic similarity ~\citep{PLACEHOLDER_bm25,PLACEHOLDER_dpr}. This practice is also enhanced with synthetic supervision from an LLM to fine-tune a model for improving retrieval performance~\citep{PLACEHOLDER_promptagator,PLACEHOLDER_inpars,PLACEHOLDER_hyde,PLACEHOLDER_synthdata_retrieval}.


We follow this paradigm and conduct a systematic empirical study of skill retrieval over a repository of \skillpool{} skills~\citep{liu2026skillusage_skillswild}. We use tasks from SkillsBench and Terminal-Bench~2 and execute approximately 1K trials across 109 tasks using hybrid retrieval pipelines in Harbor environment~\citep{Harbor_Framework}. We further construct two synthetic training tracks: Track~A ($\sim$15K pairs) creates anchor-skill tasks with retrieval-derived positives and hard negatives, while Track~B ($\sim$14K pairs) combines real data, paraphrase generation, and multi-skill task synthesis. Using these datasets, we evaluate compact retrieval models, such as \texttt{Qwen3-Embedding-0.6B} under varying fine-tuning regimes.\footnote{https://github.com/manulife-ai/emnlp2026/}


Our experiments show that synthetic supervision improves in-distribution retrieval performance but can also induce substantial forgetting. In the most aggressive fine-tuning configuration, OOD recall drops from $0.850$ to $0.650$, demonstrating that gains on synthetic training data do not necessarily translate to robust deployment behavior. 

This raises an important deployment question that has received limited attention: \emph{does synthetic-data fine-tuning degrade a model's ability to retrieve skills for real-world or out-of-distribution (OOD) tasks?} Catastrophic forgetting is well studied in continual learning, where approaches such as Learning without Forgetting (LwF) and Elastic Weight Consolidation (EWC) preserve pretrained behaviour~\citep{PLACEHOLDER_lwf,PLACEHOLDER_ewc}. However, it remains unclear how these preservation strategies behave in the context of synthetic-data-driven skill retrieval for LLM agents, especially at realistic deployment scale.

Therefore, we study several forgetting mitigation regularization approaches, including embedding-anchor regularization, LwF, EWC, and L2-initialization, under a common fine-tuning recipe \citep{PLACEHOLDER_lwf, PLACEHOLDER_ewc, li2018explicit}. Our results show that these fine-tuning regularization approaches not only retain the behavior of the frozen model on OOD or real tasks for LoRA adapters~\citep{PLACEHOLDER_lora}, but also improve performance on the synthetic (in-distribution) data compared to aggressive LoRA fine-tuning.

Our contributions are threefold. (A) We present a systematic study of skill retrieval for agents at a realistic scale, covering \skillpool{} skills and 109 verifier-scored tasks, and identify synthetic-data-induced forgetting as a deployment failure mode. (B) We introduce a unified benchmark that combines real and synthetic supervision under a shared evaluation protocol. (C)  We evaluate multiple forgetting-mitigation approaches, including embedding anchors, LwF, EWC, and L2-initialization, and show that conservative regularized fine-tuning preserves real-world and out-of-distribution behavior while improving in-distribution gains, providing a practical recipe for deployed skill-retrieval systems.



\section{Methodology}
\label{sec:methods}
This section describes the skills catalog with a retrieval server (\S\ref{sec:methods-setting}), tasks benchmarks with skills execution (\S \ref{sec:methods-realdata}), generation of synthetic skills and tasks for two tracks A (\S \ref{sec:methods-tracka}) and B (\S\ref{sec:methods-trackb}), and finally training methodology for an embedding model and a re-ranker model (\S\ref{sec:methods-training}).

\subsection{Skill Retrieval}
\label{sec:methods-setting} 
We used the skill catalogue of \skillpool{} skills collected from skillhub.club and skills.sh by \citet{liu2026skillusage_skillswild}.\footnote{www.skillhub.club, www.skills.sh} Each of these skills constitutes a folder with a SKILL.md file along with scripts and other helper files downloaded from its original GitHub repository. In addition, \citet{liu2026skillusage_skillswild} also released a retrieval server that indexed skills in two ways: (a)  skill metadata, the skill's name and description; and (b) full content in the SKILL.md file. We used both indexes to retrieve skills in different experimental settings.  The retrieval server implements BM25 retrieval, semantic retrieval using Qwen3-Embedding-4B and hybrid matching (combining BM25 and semantic matching) using reciprocal rank fusion (RRF)~\citep{PLACEHOLDER_bm25,PLACEHOLDER_qwen3embedding_rerankig,PLACEHOLDER_rrf}. 

\subsection{Real Harvest from Agent Execution}
\label{sec:methods-realdata}
We harvested real tasks and skills executions using \textsc{SkillsBench} (84 tasks) and \textsc{Terminal-Bench}~2.0 (89 tasks) benchmarks~\citep{PLACEHOLDER_skillbench,merrill2026terminalbench}. Each task provides instructions, a dockerized environment, and a verifier (test set). We execute tasks and skills in Harbor~\citep{Harbor_Framework}, a popular task execution and evaluation environment. We post-process the output of the verifier tests and convert it to a soft score $\mathrm{SR} = n_{\text{passed}} / n_{\text{total}} \in [0,1]$.

We perform task skill trials in three ways: (1) hybrid top-$k$ retrieval (full skill content) + LLM selection, (2) hybrid top-$k$ retrieval (metadata) + LLM selection, and (3) injection of ground-truth skills.  This yields $1{,}423$ trials on 173 tasks, of which $1{,}116$ produce scores (307 failed attempts). Among these, $861$ trials contain injected skills, while the rest have no skills selected by LLM.  We aggregate these trials at the $(\text{task}, \text{skills})$ tuple level by averaging $\mathrm{SR}$ across the trials, labeling a pair as (strict) positive if the mean SR exceeds $0.5$. This produces 109 tasks with reward signal (from $861$ trials), including 75 tasks with at least one positive skill, totaling 273 positive skills (3.64 per task). We use the $0.5$ threshold to favour consistently useful skills; varying it in $[0.4, 0.7]$ reduces data size but does not change baseline rankings. This 75-task / 273-pair set forms our supervised learning benchmark.

\subsection{Track A: Anchor-skill driven Synthetic Tasks}
\label{sec:methods-tracka}

We sample 1{,}772 \emph{anchor skills} from the entire skill pool using stratified sampling across 101 groups with proportional allocation and a minimum per-group of 20 (capped at 80). The groups include 51 skill categories, but only 22\% of skills have categories; for the remaining 78\%, we create 50 MiniBatch-KMeans clusters using MiniLM embeddings~\cite{wenhui_minilm}.

For each anchor skill, GPT-5 generates a self-contained task description (200–600 words) from a 3-shot prompt built from real \texttt{(skill, task)} pairs~\citep{PLACEHOLDER_gpt5}. To prevent leakage, we reject generations whose first 200 characters have Jaccard overlap $\geq 0.70$ with any real instruction and regenerate up to three times; 103 anchors are dropped, leaving 1{,}669 synthetic tasks.

For each task, we form a set of 9-skill candidates using hybrid retrieval over the full content: (i) anchor skill as a single \emph{gold positive}; (ii) four \emph{near} candidates (ranks 2--5); and (iii) four \emph{distractors} (ranks 11--50) from different \texttt{group\_label}s. Each candidate is assigned a \texttt{soft\_reward} sampled from bucket-conditioned $\mathrm{Beta}(\alpha,\beta)$ priors fitted on real data: gold $\mathrm{Beta}(0.74,0.23)$, near $\mathrm{Beta}(4,3)$, distractor $\mathrm{Beta}(1,9)$. Providing a candidate set of 14{,}687 task-skill pairs (duplicates discarded).

A candidate skill is treated as a training positive if its \texttt{soft\_reward} exceeds $0.5$, independent of its bucket of origin. This thresholding yields 5{,}463 positive task-skill pairs: 1{,}331 of 1{,}669 gold candidates (80\%), 4{,}122 of 6{,}342 near candidates (65\%), and 10 of 6{,}676 distractors (0.15\%, consistent with the $\mathrm{Beta}(1,9)$ right tail $P(X{>}0.5){=}0.5^9{\approx}0.2\%$). These 5{,}463 positive pairs involve 4{,}486 unique skills, disjoint from the 75-task real strict-positive evaluation pool (273 real positive pairs).


\subsection{Track~B: Tiered Real positives with Multi-positive Synthetic Supervision}
\label{sec:methods-trackb}
Track~B builds a multi-positive training set by combining real
$(\text{task}, \{\text{skill}_1, \dots, \text{skill}_n\})$ rows from the trial
harvest using the following method.

\textit{First}, real positives are split into two tiers: Tier-A skills have $\mathrm{soft\_reward} \geq 0.9$ and appear in $\geq 2$ trials, while Tier-B skills have $\mathrm{soft\_reward} > 0.5$. This translates to weights 1.0 and 0.7, used as weighted contribution with the loss function later.


\textit{Second}, synthetic positive rows are constructed in two ways: (i) a paraphrase process produces~$\sim20$ alternate phrasings per real task, each inheriting its positive skill set and \texttt{soft\_reward}; and (ii) a skill-first process samples $1$--$4$ \texttt{co\_used} skills and generates a task requiring them, producing intrinsically multi-positive rows. For these, \texttt{calibrated\_soft\_reward} is drawn from $\mathrm{Beta}(\mu c,(1{-}\mu)c)$, where $\mu$ is the harmonic mean of historical per-skill rewards and $c$ is the median empirical concentration (default $c=18$; 
Appx.~\ref{sec:appendix-gates}). 

\textit{Third}, each positive skill is paired with $7$ negatives from four sources: three from the top-50 semantic retriever (Qwen3-Embedding-4B), two semantically similar but unused skills ($\cos \geq 0.40$), one from BM25 top-50, and one random skill. 

\textit{Fourth}, synthetic rows are validated by a primary LLM judge (GPT-5.5) using a four-axis rubric (realism, necessity, sufficiency, reward plausibility), retaining rows with realism $\geq 4$ and others $\geq 3$. This removes $0/1{,}200$ paraphrase rows and $364/5{,}074$ skill-first rows ($7.2\%$). A second judge (Claude Opus 4.8) audits 300 samples (Gate~9); full quality gates are in Appx.~\ref{sec:appendix-gates}. After filtering, the data contains $13{,}271$ training rows ($157$ real, $4{,}256$ paraphrase, $8{,}858$ skill-first), a $556$-row validation set ($20$ real), and a $78$-row real test set.

\subsection{Training Across the Retrieval Stack}
\label{sec:methods-training}

We train models at two positions in the retrieval stack on Track-A
and Track-B (\S \ref{sec:methods-tracka}-\ref{sec:methods-trackb}): a \textbf{bi-encoder retriever}, and a
\textbf{cross-encoder reranker} using  LoRA fine-tuning and also discuss mitigation for catastrophic-forgetting signature~\citep{PLACEHOLDER_lora,PLACEHOLDER_ewc,PLACEHOLDER_lwf}


\subsubsection{Bi-encoder Retriever}
\label{sec:methods-biencoder}

Our retriever is \texttt{Qwen3-Embedding-0.6B} for which we fine-tuned LoRA adapters using InfoNCE on (tasks, positive skills, negative skills) tuples defined by
Track~A (\S\ref{sec:methods-tracka}) and Track~B (\S\ref{sec:methods-trackb})~\citep{PLACEHOLDER_qwen3embedding_rerankig,PLACEHOLDER_infonce}. For Track~B, each positive additionally carries a scalar weight $w$ (Appx.~\ref{sec:appendix-gates}) that scales its contribution to InfoNCE loss.  Together, the two tracks are complementary; they  cover diverse supervision structures (single- vs.\ multi-positive), diverse evaluation regimes (cross-validated real pool vs.\
disjoint test sets), and different LoRA adapter regimes, giving a broader view of when synthetic supervision helps or hurts than either protocol alone.

\subsubsection{Forgetting-mitigation}
\label{sec:methods-anchor}

We find that naive LoRA fine-tuning of the bi-encoder results in catastrophic forgetting (\S\ref{sec:results}).  Thus, we consider four regularization approaches. \paragraph{Embedding anchor regularization.} We regularize the embeddings of in-batch positive skills by penalizing deviations from a frozen encoder: \[ \mathcal{L}_{\mathrm{anchor}}(\theta) = \frac{1}{|\mathcal{B}^{+}|} \sum_{s \in \mathcal{B}^{+}} \bigl\| f_{\theta}(s) - f_{\theta_{0}}(s) \bigr\|_{2}^{2}, \] 

Here $f_{\theta}$ and $f_{\theta_{0}}$ the
fine-tuned and frozen skill encoders and $\mathcal{B}^{+}$ the in-batch positives. We optimize $\mathcal{L}_{\mathrm{InfoNCE}}+\lambda\mathcal{L}_{\mathrm{anchor}}$ for different $\lambda$ values. 

\paragraph{L2-init.} L2-init constrains the LoRA parameters to remain close to initialization, reducing parameter drift while allowing adaptation~\cite{li2018explicit}: \[ \mathcal{L}_{\text{L2-init}} = \frac{1}{P}\sum_{i=1}^{P} (\theta_i-\theta_i^{*})^2, \] where $\theta_i^{*}$ is the initial parameter value and $P$ is the number of trainable parameters. 

\paragraph{EWC.} Elastic Weight Consolidation (EWC) extends L2-init by weighting parameters according to their estimated importance, using a diagonal Fisher approximation computed on an anchor pool~\citep{PLACEHOLDER_ewc}: \[ \mathcal{L}_{\text{EWC}} = \frac{1}{P}\sum_{i=1}^{P} F_i(\theta_i-\theta_i^{*})^2, \] where $F_i$ is the Fisher importance of parameter $\theta_i$. 

\paragraph{LwF (KL).} Learning without Forgetting (LwF) preserves the frozen model's similarity distribution over the positive and sampled negatives using a 
temp-scaled KL divergence~\citep{PLACEHOLDER_lwf}: \[ \mathcal{L}_{\text{LwF}} = T^2\,\mathrm{KL}(p_t \,\|\, p_s). \] Here, $p_t=\mathrm{softmax}(s^{\text{teacher}}/T)$ and $p_s=\mathrm{softmax}(s^{\text{student}}/T)$ are computed from teacher and student similarity scores with $T=2.0$. Unlike parameter-based regularizers, LwF preserves ranking behaviour in the output space.

\subsubsection{Cross-encoder Reranker}
\label{sec:methods-reranker}

To evaluate whether synthetic supervision transfers from bi-encoder retrieval to point-wise relevance scoring, we fine-tune \texttt{Qwen3-Reranker-0.6B} with LoRA adapters using a listwise ranking loss \citep{cao2007learning}.

Each training instance consists of task-led query that retrieves a top-30 candidate pool, retains all positives, and fills the remaining slots with uniformly sampled negatives up to a 20-item training list. Skills are retrieved via the embedding model fine-tuned earlier. The reranker model is optimized with listwise softmax cross-entropy.  

In addition to listwise loss, we consider a listwise Learning-without-Forgetting (LwF) regularizer for the reranker.  The reranker outputs a score vector over the candidate list rather than reusable embeddings, we regularize its \emph{output distribution} instead of its parameter or embedding space. This adds a KL term between the frozen teacher's and tuned student's temperature-softened listwise score distributions, $\mathrm{KL}(\sigma(s^{\mathrm{frozen}}/T)\,\|\,\sigma(s^{\mathrm{tuned}}/T))$, with $T=2$ and $\lambda=0.1$.

We further investigate the combination of  retriever and re-ranker scores. We blend two scores after per-query min-max normalization, which rescales retriever and reranker scores independently to $[0,1]$ within each query's candidate set so they are comparable despite different raw scales; the final score is $s=\alpha s_{\text{retr}} + (1-\alpha)s_{\text{rerank}}$. Finally, we incorporate BM25 using Reciprocal Rank Fusion (RRF), which combines rankings by summing reciprocal rank contributions from BM25, retriever, and reranker pipeline. This allows lexical and semantic retrieval signals to be integrated without score calibration.

\section{Experiments}
\label{sec:experiments}

In this section, we summarize the implementation, data splits, evaluation protocol, baselines, and metrics used to assess all fine-tuned variants.

\paragraph{Implementation.} The skill index (\S\ref{sec:methods-setting}) is served via an HTTP retrieval server that supports BM25, semantic retrieval, and hybrid retrieval through RRF; all LoRA models additionally use a FAISS Flat-IP index (i.e., semantic retrieval only) on all \skillpool{} skills~\citep{PLACEHOLDER_faiss}. Real-trial harvest (\S\ref{sec:methods-realdata}) runs in the Harbor environment. We define each run as a \emph{trial}: a \texttt{(task, injected\_skills, agent)} execution with test verification, where verifiers output CTRF reports converted into  \texttt{binary reward} and \texttt{soft reward}  $\mathrm{SR} = n_{\text{passed}} / n_{\text{total}} \in [0,1]$. Track-A synthesis uses Azure-hosted GPT-5. Track-B is generated via NVIDIA Data Designer~\citep{nemo-data-designer}. Track B uses Claude~Opus~4.7 as the generator, GPT-5.5 as a judge, and Claude~Opus~4.8 for auditing 300 samples (Appx.~\ref{sec:appendix-gates})~\footnote{Note that the purpose here is not to compare two tracks but to experiment with diverse combinations of models and data generation methods.}. All models are fine-tuned using HuggingFace PEFT LoRA with AdamW on a single H100 GPU, with checkpoints selected on validation data~\citep{PLACEHOLDER_lora}.

\paragraph{Splits and evaluation.} \textbf{Track A} is evaluated on the $75$-task real strict-positive pool ($273$ pairs) under three settings:  
\textbf{(A) Synth-only:} trained on $1{,}669$ synthetic tasks and evaluated once on all real tasks (no split due to disjoint skill spaces).  
\textbf{(B) Real-only:} $5$-fold task-disjoint cross-validation over the $75$ tasks (${\sim}15$ per fold).  
\textbf{(C) Real+synth:} identical to (B) with synthetic data added to training folds; the (B)–(C) gap isolates the benefit of synthetic data.  

\textbf{Track B} uses a fixed split: $13{,}271$ training examples with $157$ real from \textsc{SkillsBench}, $556$ synthetic validation, $78$ real test samples from \textsc{SkillsBench} and \textsc{TerminalBench 2}, and 2414 synthetic task queries for testing. We train multiple retrievers by varying the LoRA recipe and synthetic-data mixture (Config~A: keeps the full mix, Config~B: drops Tier-A skill-first synth, and Config~C: keeps only skill-first; see \S\ref{sec:methods-trackb}). Each model is evaluated on three disjoint test sets:  
\textbf{Ring~1:} \textsc{SkillsBench} (real $n{=}21$),  
\textbf{Ring~2:} synthetic held-out skills ($n{=}2{,}414$),  
\textbf{Ring~3:} \textsc{TerminalBench~2} (real, $n{=}10$).  

We report \emph{effect sizes and significance tests} for all comparisons, as real data splits are very small. For small rings, we use the $10{,}000$-iteration paired bootstrap to compute $95\%$ CIs on model differences, cancelling task-level variance~\citep{PLACEHOLDER_paired_bootstrap}. For Track A, we report fold means under shared seeds so phase differences (e.g., real-only vs real+synth) are comparable.

\paragraph{Baselines.}
We compare against four frozen systems:  
(i) Hybrid retriever: \texttt{Qwen3-Embedding-4B} with BM25+RRF over \skillpool{}; (ii) Dense 0.6B retriever: \texttt{Qwen3-Embedding-0.6B}; (iii) 0.6B reranker stack: \texttt{Qwen3-Reranker-0.6B} applied to top-30 candidates; and (iv) BM25. 

\paragraph{Metrics.}
Given gold positives $G$ and top-$k$ predictions $\hat{R}_k$:  
\hitk{$k$}$=\mathbf{1}[\hat{R}_k \cap G \neq \emptyset]$,  
\recallk{$k$}$=|\hat{R}_k \cap G|/|G|$,  
\mrr$=\mathbb{E}[1/r^*]$, where $r^*$ is the rank of the first correct match; all metrics are averaged over tasks.

\section{Results}
\label{sec:results}
We report the results for our main experiments in Tables \ref{tab:tracka},  \ref{tab:rings}, and \ref{tab:p7-forget-recall10}. All metrics and the effect-size protocol are those of \S \ref{sec:experiments}.

\subsection{Bi-encoder retriever}
\label{sec:results-biencoder}

\paragraph{Track-A: synthetic supervision regresses, real-only ties.} 

Table~\ref{tab:tracka} compares Track‑A settings on the 75‑task pool. The frozen 0.6B retriever matches the much larger hybrid baseline (\hitk{10}=0.907, \recallk{10}=0.604), showing strong efficiency. Synth‑only training degrades performance (\hitk{10} -0.040, \recallk{10} -0.026). Real‑only training matches or slightly improves the baseline (\recallk{10} +0.003), indicating effective learning from real data alone.
Adding synthetic data (Real+synth) does not help: it ties Hit@10 but reduces Recall@10 by 0.021 versus Real‑only, isolating a negative synthetic contribution. The unchanged Hit@10 but lower Recall@10 suggests partial rank reshuffling—at least one correct skill remains in top‑10, but additional valid skills are pushed out. Overall, synthetic data harms multi-positive retrieval quality at this scale in Track A.


\begin{table}[t]
\centering
\small
\setlength{\tabcolsep}{4pt}
\resizebox{\columnwidth}{!}{%
\begin{tabular}{lcc}
\toprule
Setting on $75$-task pool & \hitk{10} & \recallk{10} \\
\midrule
Baseline (i): frozen $4$B + BM25 + RRF       & $0.907$ & $0.604$ \\
Baseline (ii): frozen $0.6$B (reference)     & $0.907$ & $0.604$ \\
\midrule
(A) Synth-only                                & $0.867$ & $0.578$ \\
(B) Real-only, $5$-fold task-disjoint CV     & $0.907$ & $0.607$ \\
(C) Real+synth, $5$-fold  splits like (B)    & $0.907$ & $0.586$ \\
\bottomrule
\end{tabular}%
}
\caption{Track-A bi-encoder LoRA (r=16 $\alpha$=32) trained on real and synthetic data with evaluation on the $75$-task strict-positive pool ($273$ pairs).   }
\label{tab:tracka}
\end{table}

\paragraph{Track-B: Synthetic distribution regresses again.}
\label{sec:results-finetune}

Table~\ref{tab:rings} shows the results of the multi-positive Track-B mixture and the evaluation of three rings \S \ref{sec:experiments}. The aggressive recipe of rank $32$ and rank $16$ (trained on Config~A-full-mix only) overfits the synthetic ring (Ring~2) \vs frozen $0.6$B but \emph{collapses} on the real dataset (Ring~1 and Ring~3). Table~\ref{tab:rings} also shows the results for BM25 and frozen $4$B, BM25 is the worst of all. The overfitting of aggressive LoRA is the textbook sign of catastrophic forgetting and is evident in Track A (Table~\ref{tab:tracka}). This motivates the use of forgetting mitigation regularization during LoRA training. In subsequent experiments, we focus on Track B for simplicity, as both tracks exhibit a similar regression on synthetic data. We treat the real datasets in Rings 1 and 3 as out-of-distribution (OOD), since they are either excluded from fine-tuning or constitute only a small fraction of the predominantly synthetic training data.

\begin{table}[t]
\centering
\small
\resizebox{\columnwidth}{!}{%
\begin{tabular}{lccc}
\toprule
Variant & Ring~1 & Ring~2 & Ring~3 \\
        & ($n\!=\!21$) & ($n\!=\!2414$) & ($n\!=\!10$) \\
\midrule
\multicolumn{4}{l}{\emph{Baseline: Qwen3 embedding models and BM25}} \\
frozen 0.6B          & 0.5486 & 0.5337 & 0.8500 \\
frozen 4B+BM25+RRF   & 0.5320 & 0.4609 & 0.9000 \\
BM25                 & 0.3978 & 0.1597 & 0.3500 \\
\midrule
\multicolumn{4}{l}{\emph{aggressive (all-proj, lr=2e-5)}} \\
\quad Aggr.\ ($r\!=\!16$, $\alpha\!=\!32$) & 0.507 & 0.577 & 0.650 \\
\quad Aggr.\ ($r\!=\!32$, $\alpha\!=\!64$) & 0.498 & 0.566 & 0.700 \\
\bottomrule
\end{tabular}
}
\caption{\recallk{10} for BM25, baseline embedding models and aggressive LoRA on the three evaluation rings using Track B's synthetic data, Config~A.}
\label{tab:rings}
\end{table}

\paragraph{Forgetting mitigation approaches.}
\label{sec:results-track-B}

Table~\ref{tab:p7-forget-recall10} compares four forgetting-mitigation approaches (see \S\ref{sec:methods-anchor}): conservative anchor, LwF, L2-init, and EWC. Results are reported on the three Rings for three Track B configurations: A (full mix), B (skills first removed), and C (only skills first). We find that Recall@10 remained similar for all approaches. If we compare with the baselines and aggressive LoRA fine-tuning in Table~\ref{tab:rings} (see Config A), these approaches achieve similar or better performance on all evaluation Rings. Table~\ref{tab:p7-forget-recall10} also shows that the differences between forgetting-mitigation methods are negligible, while all methods improve recall on the in-distribution synthetic data of Ring 2. On the OOD datasets of Rings 1 and 3, they maintain recall comparable to the baseline embedding models. Thus forgetting-mitigation methods not only prevent forgetting but also improve performance on the new training distribution.


We further swept anchor weight ($\lambda$), LoRA rank ($r$), and random seeds for the conservative anchor method (see Appx.~\ref{sec:appendix-lambda-seed}). The results show that LoRA rank $r$ is the primary factor controlling the performance-retention trade-off, whereas anchor weight ($\lambda$) has only a minor effect. Low-rank conservative updates ($r=8$) exhibit minimal seed sensitivity and consistently preserve real/OOD performance while retaining most gains on synthetic data. In contrast, unregularized aggressive updates show substantially higher variance across seeds, indicating reduced training stability.

\begin{table}[t]
\centering
\setlength{\tabcolsep}{4pt}
\resizebox{\columnwidth}{!}{%
\begin{tabular}{llccc}
\toprule
Config & Method & Ring~1 & Ring~2 & Ring~3 \\
 & & \scriptsize $n{=}21$ & \scriptsize $n{=}2414$ & \scriptsize $n{=}10$ \\
\midrule
\multirow{4}{*}{Config A} & Cons.\ anchor( $\lambda\!=\!0.1$) & 0.540 & 0.6083 & 0.8500 \\
 & LwF          & 0.540 & 0.6077 & 0.8500 \\
 & EWC          & 0.540 & 0.6067 & 0.8500 \\
 & L2$\to$init  & 0.540 & 0.6076 & 0.8500 \\
\midrule
\multirow{4}{*}{Config B} & Cons.\ anchor ($\lambda\!=\!0.1$)& 0.5554 & 0.5685 & 0.8500 \\
 & LwF          & 0.5554 & 0.5610 & 0.8500 \\
 & EWC          & 0.5554 & 0.5606 & 0.8500 \\
 & L2$\to$init  & 0.5486 & 0.5638 & 0.8500 \\
\midrule
\multirow{4}{*}{Config C} & Cons.\ anchor ($\lambda\!=\!0.1$) & 0.5463 & 0.5626 & 0.8500 \\
 & LwF          & 0.5463 & 0.5703 & 0.8500 \\
 & EWC          & 0.5463 & 0.5704 & 0.8500 \\
 & L2$\to$init  & 0.5463 & 0.5657 & 0.8500 \\

\bottomrule
\end{tabular}%
}
\caption{Recall@10  for four forgetting mitigation approaches on  Track B's synthetic data configurations (LoRA: r=8, $\alpha$=16 attn-only, lr=5e-6). }
\label{tab:p7-forget-recall10}
\end{table}

\paragraph{Effect Size and Large OOD dataset.}

To assess practical significance beyond aggregate metrics, we performed a paired per-task analysis against the frozen-0.6B baseline (Appx.~\ref{appendix:effect_size}). Across both Ring~1 ($N=21$) and Ring~3 ($N=10$), all forgetting-mitigation methods (Conservative Anchor, LwF, EWC, and L2-init) remained effectively unchanged relative to the baseline, with near-zero mean differences ($\Delta=-0.009$ on Ring~1, $\Delta=0.0$ on Ring~3), negligible effect sizes, and TOST confirming practical equivalence. 

\begin{table*}[t]
\centering
\scriptsize
\setlength{\tabcolsep}{4pt}
\resizebox{\textwidth}{!}{%
\begin{tabular}{p{6.8cm}c|ccc|ccc}
\toprule
& & \multicolumn{3}{c|}{\textbf{Real 75 tasks (out-of-distribution)}} & \multicolumn{3}{c}{\textbf{Synth 1,056 tasks (in-distribution)}} \\
Method & $\alpha$ & h@1 & h@5 & MRR & h@1 & h@5 & MRR \\
\midrule
Retriever only & 1.00 & 0.640 & 0.840 & 0.735 & 0.469 & 0.694 & 0.582 \\
Frozen reranker only & 0.00 & 0.653 & 0.813 & 0.721 & 0.566 & 0.813 & 0.677 \\
Frozen reranker + retriever blend & 0.55 & 0.693 & 0.880 & 0.773 & 0.566 & 0.813 & 0.677 \\
LoRA reranker only & 0.00 & 0.653 & 0.853 & 0.735 & 0.617 & 0.834 & 0.712 \\
LoRA reranker + retriever blend & 0.55 & \textbf{0.707} & \textbf{0.893} & \textbf{0.781} & 0.617 & 0.834 & 0.712 \\
LoRA reranker with listwise LwF ($\lambda = 0.1$) & 0.00 & 0.640 & 0.853 & 0.745 & \textbf{0.653} & \textbf{0.861} & \textbf{0.744} \\
LoRA reranker with listwise LwF + retriever blend ($\lambda = 0.1$) & 0.50 & \textbf{0.707} & \textbf{0.893} & 0.780 & \textbf{0.653} & \textbf{0.861} & \textbf{0.744} \\

LoRA reranker only with listwise LwF ($\lambda = 0.3$) & 0.00 & 0.667 & 0.867 & 0.765 & 0.632 & 0.831 & 0.723 \\
LoRA reranker with listwise LwF+ retriever blend ($\lambda = 0.3$) & 0.45 & 0.693 & 0.880 & 0.771 & 0.622 & 0.835 & 0.717 \\
\bottomrule
\end{tabular}%
}
\caption{Comparison of hit@k and MRR for retriever and  reranker model with different combinations. Blended score measured as: $s = \alpha s_{\text{retr}} + (1-\alpha)s_{\text{rerank}}$.}
\label{tab:reranker_compact}
\end{table*}
In contrast, aggressive full-projection adaptation consistently produced larger negative shifts ($\Delta=-0.051$ to $-0.250$) and more task-level regressions. Notably, the aggressive $r=16,\alpha=32$ model showed statistically significant degradation on Ring~3 ($0/4/6$ wins/ties/losses, $\Delta=-0.400$, Wilcoxon/sign-test $p=0.0312$).  Overall, these results support our main conclusion that conservative forgetting-mitigation recipes preserve prior capabilities, whereas aggressive adaptation induces a higher risk of regression on OOD tasks. Full per-task results, effect sizes, confidence intervals, and significance tests are reported in Appx.~\ref{appendix:effect_size}.

To complement the small-sample real-task analysis in Appx.~\ref{appendix:effect_size}, we evaluated all methods on six BEIR datasets ~\cite{BEIR_etal}. BEIR benchmark provides a larger and more diverse OOD retrieval benchmark (Appx.~\ref{appendix:large_sample_size}). Across the six BEIR datasets, the forgetting-mitigated approaches 
preserved or slightly improved Recall@10 and MRR@10 relative to the Frozen-0.6B baseline, while they already showed higher in-distribution performance in Table~\ref{tab:rings} and Table~\ref{tab:p7-forget-recall10}. In contrast, aggressive LoRA fine-tuning consistently reduced performance, often by a substantial margin. These results mirror our findings in Table~\ref{tab:rings} and Table~\ref{tab:p7-forget-recall10} that conservative forgetting-mitigation preserves retrieval quality on OOD data while improving on in-distribution data; and aggressive adaptation harms both in-distribution and OOD generalization. Full results are in Appx.~\ref{appendix:large_sample_size}.




\subsection{Cross-encoder reranker }
\label{sec:results-rerank}



In this section, we evaluate reranker \texttt{Qwen3-Reranker-0.6B} on top of the conservative anchor ($\lambda=0.1$) regularized fine-tuned \texttt{Qwen3-Embedding-0.6B} retriever. We fine-tune \texttt{Qwen3-Reranker-0.6B} (as explained in \S \ref{sec:methods-reranker}) on 1{,}669 synthetic tasks from Config A of Track B, and evaluate  on 75 real tasks from \textsc{SkillsBench} and \textsc{TerminalBench~2}, and synthetic data split of Track B's configuration A. Table~\ref{tab:reranker_compact} shows the results. On the 75 real deployment tasks, the retriever alone reaches $h@5=0.840$ and $\mathrm{MRR}=0.735$. When we blend it with a frozen \texttt{Qwen3-Reranker-0.6B} improves this to $h@5=0.880$ and $\mathrm{MRR}=0.773$ at $\alpha=0.55$ (blending is explained in \S \ref{sec:methods-reranker}). When reranker  is fine-tuned with LoRA on top-30 candidate lists, it reaches the real-pool ceiling at $h@5=0.893$ with $\mathrm{MRR}=0.781$ with a retriever blend. Table~\ref{tab:reranker_compact} also shows that the LwF regularizer based reranker fine-tuning (see \S \ref{sec:methods-reranker}) is neutral on the 75-task real pool, where it also reaches $h@5=0.893$, but it gives the strongest results on the in-distribution synthetic split $h@5=0.861$ and $\mathrm{MRR}=0.744$. This observation is the same as experiments with retrieval model where we found that all of the forgetting mitigation regularizers  result in improved performance on in-distribution data and showed the same or better performance on out-of-distribution than the baseline model.



In addition, we also experimented with adding of BM25 through reciprocal-rank fusion, and it does not improve the reranker ceiling.   So, our final stack uses the fine-tuned LoRA anchor regularized Qwen3-Embedding-0.6B and LoRA regularized Qwen3-Reranker-0.6B with blended scoring.



\section{Related Work}
\label{sec:related}

\paragraph{Skills, retrieval, and synthetic supervision for agents.}
Agent systems increasingly rely on retrieved procedures, tools, or skills at run time~\citep{PLACEHOLDER_anthropic_skills,PLACEHOLDER_react}, and generating queries or pseudo-labels with LLMs is a standard recipe for low-supervision retrieval~\citep{nogueira2020passagereranking,PLACEHOLDER_promptagator,PLACEHOLDER_inpars,PLACEHOLDER_hyde,PLACEHOLDER_synthdata_retrieval}. While previous work on skill router training reports gains from synthetic data \citep{zheng2026skillrouter}, our findings show a nuanced picture: synthetic supervision is beneficial when applied to lightweight components but can harm performance when used to fine-tune encoders or rerankers if the synthetic task fails to capture the true multi-positive structure of downstream evaluation. 


\paragraph{Continual learning and catastrophic forgetting.}
Catastrophic forgetting has been widely studied in continual learning methods, such as: EWC regularizes parameters against the Fisher-information-weighted posterior of the previous task~\citep{PLACEHOLDER_ewc}, learning without forgetting (LwF) regularizes outputs via knowledge distillation~\citep{PLACEHOLDER_lwf}, L2-init keeps parameters close to initialization \citep{li2018explicit}, and analyzing the extent of catastrophic forgetting in neural ranking models~\citep{lovon2021studying}. We present an empirical study of synthetic data induced forgetting in large-scale skill retrieval for LLM agents.  We compare preservation-style regularization--including LwF, EWC, L2-init, and embedding anchoring--under a shared conservative LoRA recipe.


\section{Conclusion}

We studied skill retrieval for LLM agents on the \skillpool{}-skill scale using a unified benchmark that combines real tasks and synthetic supervision. Our experiments show that naive fine-tuning can degrade real-task retrieval and induce catastrophic forgetting on out-of-distribution queries. Our results also show that the use of forgetting mitigation regularizers with LoRA based fine-tuning not only improves in-distribution performance but also preserves the out-of-distribution performance. According to our intervention-based experiments, performance regression in aggressive updates can be characterized as stability-plasticity failure consistent with representational or ranking-function drift. Additionally, we show that compact retrievers (0.6B) can match much larger hybrid systems, emphasizing the importance of supervision quality over model scale.


\section*{Limitations}

\paragraph{Statistical power.}

Val-set~1 ($n\!=\!21$) and Val-set~3 ($n\!=\!10$) are limited in size, such that single-task variations can shift \recallk{10} by approximately $5\%$. We mitigate this by using 10,000-iteration paired bootstrap sampling, detailed statistical analysis on effect size and significance (see Appx. \ref{appendix:effect_size}), and evaluation on large out-of-distribution BEIR benchmark (Appx. \ref{appendix:large_sample_size}). 

\paragraph{Known quality-gate failures in Track B.}
Although Track~B applies multiple quality gates, three important limitations remain. First, the synthetic and real \texttt{soft\_reward} distributions differ substantially. Second, positive-skill coverage is sparse relative to the full catalogue. Third, inter-judge agreement between the two LLM evaluators is modest. We report these issues explicitly because they affect how strongly the synthetic mixture can be treated as a faithful proxy for real supervision. Detailed mitigation on soft\_reward and  inter-judge agreement is discussed in Appendix \ref{sec:appendix-gates}. Overall, the paper mitigates the impact by showing detailed empirical results on forgetting mitigation approaches and their positive impact on LoRA fine-tuning when generating synthetic data for sparse positive data is difficult.

\paragraph{Scale and coverage of the training signal.}
Our experiments are conducted on a skill pool of approximately 34K items, but the number of positively supervised skills is still a small fraction of the full catalogue. We therefore cannot rule out the possibility that some negative results--especially for encoder LoRA fine-tuning--would change at substantially larger synthetic-data scale or under broader positive-skill coverage. We do mitigate this concern by showing results on external large benchmarks and how forgetting mitigation methods can preserve the existing behaviour while improving results on a new training distribution.

\paragraph{Scope of domains, language, and modality.}
In adherence to the Bender Rule \cite{bender2019benderrule}, we note that all tasks we executed in our experiments are English-language, text-based, and evaluated through a dockerized execution harness. We do not study multilingual skill retrieval, multimodal inputs, or non-code agent settings. The extent to which the observed forgetting patterns and mitigation approaches transfer to other domains remains an open question.

\paragraph{Dependence on closed-weight LLMs.}
Both synthetic pipelines rely on commercial, closed-weight LLMs for generation and quality assessment. As a result, exact reproduction of the synthetic data and judge labels may be sensitive to model or API drift over time. While we document the pipeline in detail, full reproducibility is therefore constrained by external endpoints.

\section*{Ethics Statement}

The skill catalogue and benchmark tasks are operated under license from their respective owners; we add no personally identifying information at any stage. LLM-generated synthetic tasks are decontaminated against the benchmark instructions by a Jaccard reject filter (max observed overlap on the spot check: $0.129$); generated tasks describe generic engineering problems and were spot-checked for offensive content. Closed-weight LLMs are used both as generators and as judges; we deliberately split vendors between generator and judge to reduce self-confirmation bias and report inter-vendor agreement as a quality gate.  Human subjects randomly sampled the synthetic data to assess its quality. AI tools were also used to assist authors in polishing writing and generating code, and all the ideas and reasoning are original from the authors. 

\bibliography{custom}

\appendix



\section{Track-A Generalization Safeguards}
\label{sec:appendix-safeguards}

Three independent leakage probes are run on every Track-A experiment:
\begin{itemize}
  \setlength{\itemsep}{1pt}
  \item \emph{Task disjointness}: the 1{,}669 synthetic tasks share 0 \texttt{task\_id} with the 109 real tasks (verified at every run).
  \item \emph{Skill disjointness}: the 9{,}254 distinct synth-positive skills do not overlap with the 332 distinct real-positive skills. Every synth-only experiment is therefore a strict cross-skill-subspace generalization test.
  \item \emph{Decontamination spot check}: 50 random synthetic tasks vs all 176 real benchmark instructions; max observed Jaccard@200 chars $=0.129$, no instance flagged at $0.50$.

\end{itemize}

\section{Track B Details and Quality Gates}
\label{sec:appendix-gates}

\begin{table}[h]
\centering\small
\begin{tabular}{@{}clp{4.0cm}c@{}}
\toprule
\# & Type & Check & Status \\
\midrule
G1  & schema   & all fields present, $|N|{=}7$, no positive overlap & pass \\
G2  & coverage & every train task has $\geq 1$ Tier-A/B real positive & pass \\
G3  & judge    & paraphrase retention $\geq 80\%$ per source task & pass \\
G4  & judge    & negative-leakage audit ($n{=}100$) & pass \\
G5  & embed.   & frozen-encoder collapse check (smoke fine-tune) & pass \\
G6  & leakage  & no train/val text or seed set equals any test row & pass \\
G7  & dist.    & KS-test of synthetic vs.\ real \texttt{soft\_reward} & soft fail \\
G8  & coverage & fraction of pool appearing as a positive & soft fail \\
G9  & judge    & inter-judge agreement (300-row audit) & soft fail \\
G10 & holdout  & no \texttt{eval\_skill\_holdout} skill in train/val & pass \\
\bottomrule
\end{tabular}
\caption{Quality gates for the locked Track~B build. ``Soft fail'' gates are
limitation flags: they diagnose mismatch between the synthetic supervision and
the real harvest, but do not imply that the retained rows are
invalid.}
\label{tab:gates}
\end{table}

Synthetic Track~B rows pass through ten gates before joining the locked split (Table~\ref{tab:gates}).
Six are hard build checks for schema validity, leakage prevention, coverage of
real tasks, and held-out-skill integrity. Four (G3, G7--G9) are statistical or
judge-based diagnostics. G3 is a retention gate that can directly block
low-quality synthetic rows, whereas G7--G9 are reported as limitation flags
rather than build blockers: they test reward calibration, catalog coverage, and
judge calibration, not whether the retained task--skill pairs are internally
coherent. We therefore use these gates to bound the interpretation of Track~B
rather than to claim that the synthetic supervision is distributionally
equivalent to the real harvest.

\paragraph{Generator and primary judge.}
Synthetic tasks are produced by Claude~Opus~4.7. Every row is then scored
by GPT-5.5 on a four-axis 1--5 Likert rubric:
\emph{task realism--could a real user plausibly send this task?}; \emph{skill necessity--are all listed skills required, not merely helpful?}; \emph{skill sufficiency--is the listed skill set sufficient to solve the task?},
\emph{reward plausibility--does the \texttt{calibrated\_soft\_reward} feel right?}

A row passes iff $\text{realism}\!\geq\!4 \wedge
 \min(\text{necessity}, \text{sufficiency}, \text{reward\_plausibility.}) \geq 3$.
The asymmetric realism bar reflects the dataset's purpose: training value
hinges on tasks feeling like real user input. Rows are weighted by
$w = \texttt{calibrated\_soft\_reward} \times
   \min(\text{realism}, \text{necessity}, \text{sufficiency})/5$,
so marginal passes also receive lower contrastive weight. This judge rejects $0/1{,}200$ paraphrase rows and $364/5{,}074$
($7.2\%$) skill-first rows; failures are dominated by sufficiency
(mean $3.94$).

\paragraph{G7 --- reward distribution.}
A two-sample KS test comparing real-trial \texttt{soft\_reward} values and
synthetic \texttt{calibrated\_soft\_reward} values
($n_{\text{real}}{=}1{,}116$, $n_{\text{synth}}{=}8{,}858$) yields
$D{=}0.31$, $p{<}10^{-80}$, so the synthetic reward weights are not a
distributional match to the real harvest. This is a soft fail because
\texttt{calibrated\_soft\_reward} is not intended to be an empirical execution
reward; it is a synthetic confidence prior used only to weight the contrastive
loss through \texttt{positive\_weight}. For skill-first rows it is sampled from
$\mathrm{Beta}(\mu c, (1{-}\mu)c)$, where $\mu$ is derived from historical
per-skill reward means and $c$ controls concentration, so the resulting
distribution is expected to be smoother and more peaked than the long-tailed
real-trial distribution. We therefore interpret G7 as a calibration-mismatch
warning on the weighting signal, not as evidence that the retained task--skill
assignments are implausible. 

\paragraph{G8 --- skill coverage.}
Of the $26{,}947$ retrieval-eligible skills (the full $34{,}396$ pool minus the
held-out subset), only $140$ ($0.52\%$) appear as positives in the locked train
set. This is a structural limitation of Track~B rather than a failed attempt at
full-catalog supervision. The skill-first generator samples only from skills
with at least one successful real trial, and the paraphrase branch simply
reuses positives from the same small real seed set, so positive supervision is
intentionally concentrated on a narrow supported subset of the catalog.
Track~B therefore provides dense supervision over a small region of the pool,
not broad positive coverage of the long tail. We mark G8 as a soft fail because
it limits how strongly any positive result can be interpreted as full-pool
adaptation, and it motivates retaining a strong frozen-retriever fallback at
inference rather than treating the fine-tuned model as a standalone long-tail
index.

\paragraph{G9 --- inter-judge agreement.}
A second judge (Claude~Opus~4.8) re-scores a stratified $300$-row audit
($150$ rows originally passed by the primary judge and $150$ originally failed)
under the identical rubric. The means per-axis align most closely with
\emph{necessity} ($4.44$ vs.\ $4.45$, Pearson $r{=}0.79$), the axis most
directly tied to whether the listed skills are actually required. On the binary
pass decision, Cohen's $\kappa{=}0.28$ and raw agreement is $64\%$, so Gate~9
is recorded as a soft fail. However, $\kappa$ alone obscures the structure of
the disagreement. The McNemar table
$(n_{11},n_{10},n_{01},n_{00})\!=\!(144,6,102,48)$,
$p{=}1.3{\times}10^{-23}$ shows that the disagreement is strongly asymmetric:
GPT-5.5 is the stricter judge, with $144/150$ of its accepted rows also
accepted by the second judge, while many rows it rejected are accepted by
Claude~Opus~4.8. Therefore, we interpret the retained positives as a
conservative, high-precision subset rather than as an unstable label set because our main judge is GPT 5.5 and Track B kept only GPT 5.5 judged rows. This means that the evidence is more consistent with threshold mismatch
than with arbitrary labeling noise.

\paragraph{G10 — held-out-skill integrity.}
At split time we freeze ${\sim}3{,}000$ skills ($\sim$9\% of the pool) into
\texttt{eval\_skill\_holdout} — skills reserved exclusively for building the
held-out-skill synthetic eval ring later, and never allowed to touch training.
A skill can leak into a training/validation row in three different ways: (1) as the \emph{positive} — the skill labeled as the
correct answer for a query; (2) as a \emph{negative} — a skill listed as a
wrong/distractor candidate for a query; or (3) as a \emph{seed} — in our
skill-first generation track, some tasks are synthesized by starting from a
skill and asking an LLM to invent a plausible task for it, so that seed skill
must also be holdout-free. G10 counts how many rows fail each of these three
checks; G10 evaluation report shows: \textit{0 positive violations, 0 negative
violations, and 0 seed violations}. This is a build-blocking gate — a single
violation in any of the three would mean the model has already "seen" a
supposedly held-out skill during training, silently invalidating every
held-out-eval number in the paper — so its pass status is load-bearing for other experiments.

\section {Lambda Sweep and Seed Variance}
\label{sec:appendix-lambda-seed}

Table~\ref{tab:p7-lambda-rank-grid} shows the effect of LoRA rank $r$ and anchor weight $\lambda$ on the conservative forgetting mitigation recipe. The main variation comes from rank, not from $\lambda$. At fixed $\lambda$, increasing $r$ changes Recall@10 by up to $0.050$ across rings, whereas sweeping $\lambda$ at fixed $(r,\mathrm{ring})$ changes Recall@10 by at most $0.012$ and usually much less. The middle synthetic ring improves from $r=4$ to $r=16$ and then saturates, while Ring 3 remains unchanged for $r \leq 8$ but drops from $0.850$ to $0.800$ at $r \geq 16$. Ring~1 shows a slight decrease Recall@1 from r=4 to r=16, up to 0.007 then goes down aggressively by 0.021 on r=32.  Thus, in this sweep, rank is the primary capacity/retention trade-off knob, whereas the conservative anchor weight has only a small effect within a fixed rank.

\begin{table}[t]
\centering
\small
\setlength{\tabcolsep}{4pt}

\resizebox{\linewidth}{!}{%
\begin{tabular}{llccccc}
\toprule
$r$ & Ring & $\lambda{=}0$ & $\lambda{=}0.03$ & $\lambda{=}0.1$ & $\lambda{=}0.3$ & row spread \\
\midrule
\multirow{3}{*}{4} & Ring 1   & 0.546 & 0.546 & 0.546 & 0.546 & 0.000 \\
                   & Ring 2 & 0.583 & 0.583 & 0.583 & 0.586 & 0.004 \\
                   & Ring 3   & 0.850 & 0.850 & 0.850 & 0.850 & 0.000 \\
\midrule
\multirow{3}{*}{8} & Ring 1   & 0.539 & 0.539 & 0.539 & 0.539 & 0.000 \\
                   & Ring 2 & 0.610 & 0.608 & 0.610 & 0.606 & 0.004 \\
                   & Ring 3   & 0.850 & 0.850 & 0.850 & 0.850 & 0.000 \\
\midrule
\multirow{3}{*}{16} & Ring 1   & 0.535 & 0.524 & 0.524 & 0.535 & 0.012 \\
                    & Ring 2 & 0.620 & 0.619 & 0.618 & 0.620 & 0.003 \\
                    & Ring 3   & 0.800 & 0.800 & 0.800 & 0.800 & 0.000 \\
\midrule
\multirow{3}{*}{32} & Ring 1   & 0.514 & 0.514 & 0.514 & 0.514 & 0.000 \\
                    & Ring 2 & 0.615 & 0.612 & 0.611 & 0.613 & 0.003 \\
                    & Ring 3   & 0.800 & 0.800 & 0.800 & 0.800 & 0.000 \\
\midrule
\multirow{3}{*}{col spread} & Ring 1   & 0.032 & 0.032 & 0.032 & 0.032 & \\
                            & Ring 2 & 0.037 & 0.036 & 0.035 & 0.034 & \\
                            & Ring 3   & 0.050 & 0.050 & 0.050 & 0.050 & \\
\bottomrule
\end{tabular} }
\caption{Stage A: Recall@10 on the three evaluation rings for the $(r,\lambda)$ grid at seed=42 (conservative anchor regulariz: attn-only, lr=5e-6, Config A / full mix).  \emph{Row spread} is max$-$min of Recall@10 across $\lambda$ at fixed $(r,\text{ring})$; \emph{col spread} is max$-$min across $r$ at fixed $(\lambda,\text{ring})$. }
\label{tab:p7-lambda-rank-grid}
\end{table}

Table~\ref{tab:p7-seed-varaince} shows the seed sensitivity for conservative and aggressive recipes. The conservative $r=8$ runs are highly stable: the variance is effectively zero on Ring~1 and Ring~3 and remains negligible on the middle synthetic ring. The variance increases at higher rank, especially in Ring1 and Ring 3; e.g., for conservative $r=32$, the Ring~3 standard deviation increases to $0.029$ at $\lambda=0.0$ and $0.076$ at $\lambda=0.1$. The unregularized aggressive all-projection runs are both less stable and less reliable, with the largest variance again concentrated on Ring~3 ($0.115$ for $r=16$ and $0.076$ for $r=32$). Overall, these results indicate that the conservative low-rank regularized setting lies below the seed-noise floor, whereas higher-capacity and unregularized recipes are materially more seed-sensitive.
 
Overall, lower-capacity regularized updates retain real/OOD behavior while preserving most synthetic gains. The regression of performance on aggressive updates, according to our intervention-based updates, can be characterized as a stability-plasticity failure consistent with representational or ranking-function drift. 

\begin{table}[t]
\centering
\small
\setlength{\tabcolsep}{4pt}
\resizebox{\linewidth}{!}{%
\begin{tabular}{lcccc}
\toprule
Recipe  & Ring 1 & Ring 2 & Ring 3 \\
\midrule
cons $r{=}8$, $ α=16, \lambda{=}0.0$          &  $0.539 \pm 0.000$ & $0.609 \pm 0.004$ & $0.850 \pm 0.000$ \\
cons $r{=}8$, $α=16,\lambda{=}0.1$          &  $0.539 \pm 0.000$ & $0.610 \pm 0.001$ & $0.850 \pm 0.000$ \\
cons $r{=}32$, $α=64,\lambda{=}0.0$         &  $0.515 \pm 0.006$ & $0.614 \pm 0.002$ & $0.817 \pm 0.029$ \\
cons $r{=}32$, $α=64,\lambda{=}0.1$         & $0.501 \pm 0.016$ & $0.615 \pm 0.003$ & $0.783 \pm 0.076$ \\
aggressive (r=16 α=32 all-proj)  & $0.505 \pm 0.022$ & $0.586 \pm 0.026$ & $0.583 \pm 0.115$ \\
aggressive (r=32 α=64 all-proj)  & $0.492 \pm 0.009$ & $0.600 \pm 0.031$ & $0.683 \pm 0.076$ \\
\bottomrule
\end{tabular}}
\caption{Seed variance for the (r, $\lambda$) sweep. Each cell is mean $\pm$ sample std over the seeds listed. Seed set = \{42, 1337, 99991\}. Each cell shows the mean Recall@10 across three seeds and the sample standard deviation. }
\label{tab:p7-seed-varaince}
\end{table}

\section { Effect Size and Significance}
\label{appendix:effect_size}

\begin{table*}[t]
\centering
\footnotesize
\setlength{\tabcolsep}{3pt}
\resizebox{\linewidth}{!}{%
\begin{tabular}{lcccccc}
\toprule
Task & Cons & LwF (KL) & EWC & L2-init & Aggr.\ ($r\!=\!16$, $\alpha\!=\!32$)  & Aggr.\ ($r\!=\!32$, $\alpha\!=\!64$) \\
\midrule
citation-check & +0.000 & +0.000 & +0.000 & +0.000 & +0.000 & +0.000 \\
data-to-d3 & +0.000 & +0.000 & +0.000 & +0.000 & -0.333 & -0.333 \\
dynamic-object-aware-egomotion & +0.000 & +0.000 & +0.000 & +0.000 & -0.286 & -0.143 \\
earthquake-plate-calculation & +0.000 & +0.000 & +0.000 & +0.000 & +0.000 & +0.000 \\
econ-detrending-correlation & +0.000 & +0.000 & +0.000 & +0.000 & +0.000 & +0.000 \\
fix-build-google-auto & +0.000 & +0.000 & +0.000 & +0.000 & +0.000 & +0.000 \\
glm-lake-mendota & +0.000 & +0.000 & +0.000 & +0.000 & +0.000 & +0.000 \\
gravitational-wave-detection & +0.000 & +0.000 & +0.000 & +0.000 & +0.000 & +0.000 \\
hvac-control & +0.000 & +0.000 & +0.000 & +0.000 & +0.000 & +0.286 \\
manufacturing-equipment-maintenance & +0.000 & +0.000 & +0.000 & +0.000 & +0.000 & +0.000 \\
manufacturing-fjsp-optimization & +0.000 & +0.000 & +0.000 & +0.000 & -0.250 & -0.250 \\
mario-coin-counting & -0.250 & -0.250 & -0.250 & -0.250 & -0.250 & -0.500 \\
multilingual-video-dubbing & +0.000 & +0.000 & +0.000 & +0.000 & -0.143 & -0.143 \\
parallel-tfidf-search & +0.000 & +0.000 & +0.000 & +0.000 & +0.000 & +0.000 \\
pg-essay-to-audiobook & +0.000 & +0.000 & +0.000 & +0.000 & +0.500 & +0.500 \\
quantum-numerical-simulation & +0.000 & +0.000 & +0.000 & +0.000 & +0.000 & +0.000 \\
sales-pivot-analysis & +0.200 & +0.200 & +0.200 & +0.200 & +0.200 & +0.000 \\
speaker-diarization-subtitles & -0.143 & -0.143 & -0.143 & -0.143 & -0.286 & -0.286 \\
syzkaller-ppdev-syzlang & +0.000 & +0.000 & +0.000 & +0.000 & +0.000 & +0.000 \\
trend-anomaly-causal-inference & +0.000 & +0.000 & +0.000 & +0.000 & +0.000 & +0.000 \\
video-filler-word-remover & +0.000 & +0.000 & +0.000 & +0.000 & -0.400 & -0.200 \\
\midrule
\textbf{Mean $\Delta$} & -0.009 & -0.009 & -0.009 & -0.009 & -0.059 & -0.051 \\
\bottomrule
\end{tabular}}
\caption{Per-task $\Delta$Recall@10 (method $-$ Frozen 0.6B) on Ring 1 trained on Track B's Config A. Positive = method wins, negative = method regresses vs the frozen baseline.}
\label{tab:per-task_ring1}
\end{table*}

\begin{table*}[t]
\centering
\footnotesize
\setlength{\tabcolsep}{3pt}
\resizebox{\linewidth}{!}{%
\begin{tabular}{lcccccc}
\toprule
Task & Cons & LwF (KL) & EWC & L2--init & Aggr.\ ($r\!=\!16$, $\alpha\!=\!32$) & Aggr.\ ($r\!=\!32$, $\alpha\!=\!64$) \\
\midrule
adaptive-rejection-sampler & +0.000 & +0.000 & +0.000 & +0.000 & +0.000 & -0.500 \\
make-doom-for-mips & +0.000 & +0.000 & +0.000 & +0.000 & -1.000 & -1.000 \\
openssl-selfsigned-cert & +0.000 & +0.000 & +0.000 & +0.000 & +0.000 & +0.000 \\
portfolio-optimization & +0.000 & +0.000 & +0.000 & +0.000 & +0.000 & +0.000 \\
pypi-server & +0.000 & +0.000 & +0.000 & +0.000 & -0.500 & -1.000 \\
qemu-startup & +0.000 & +0.000 & +0.000 & +0.000 & +0.000 & +0.500 \\
regex-chess & +0.000 & +0.000 & +0.000 & +0.000 & +0.000 & +0.000 \\
sanitize-git-repo & +0.000 & +0.000 & +0.000 & +0.000 & +0.000 & +0.000 \\
schemelike-metacircular-eval & +0.000 & +0.000 & +0.000 & +0.000 & -0.500 & -0.500 \\
video-processing & +0.000 & +0.000 & +0.000 & +0.000 & +0.000 & +0.000 \\
\midrule
\textbf{Mean $\Delta$} & +0.000 & +0.000 & +0.000 & +0.000 & -0.200 & -0.250 \\
\bottomrule
\end{tabular}}
\caption{Per-task $\Delta$Recall@10 (method $-$ Frozen 0.6B) on Ring 3  trained on Track B's Config A(full mix). Positive = method wins, negative = method regresses vs the frozen baseline.}
\label{tab:per-task_ring3}
\end{table*}

\begin{table*}[t]
\centering
\footnotesize
\setlength{\tabcolsep}{3pt}
\resizebox{\linewidth}{!}{%
\begin{tabular}{llrrrrrrrr}
\toprule
Method & Ring & $N$ & W/T/L & mean $\Delta$ & 95\% CI & Wilcoxon $p$ & sign $p$ & Cliff's$\delta$ & TOST $p$ \\
\midrule
Conservative Anchor LoRA & Ring 1 & 21 & 1/18/2 & -0.009 & [-0.043, +0.022] & 0.75 & 1 & -0.048 & 0.013 \\
Conservative Anchor LoRA & Ring 3 & 10 & 0/10/0 & +0.000 & [+0.000, +0.000] & 1 & 1 & +0.000 & 0 \\
LwF (KL) LoRA & Ring 1 & 21 & 1/18/2 & -0.009 & [-0.043, +0.022] & 0.75 & 1 & -0.048 & 0.013 \\
LwF (KL) LoRA & Ring 3 & 10 & 0/10/0 & +0.000 & [+0.000, +0.000] & 1 & 1 & +0.000 & 0 \\
EWC LoRA & Ring 1 & 21 & 1/18/2 & -0.009 & [-0.043, +0.022] & 0.75 & 1 & -0.048 & 0.013 \\
EWC LoRA & Ring 3 & 10 & 0/10/0 & +0.000 & [+0.000, +0.000] & 1 & 1 & +0.000 & 0 \\
L2-init LoRA & Ring 1 & 21 & 1/18/2 & -0.009 & [-0.043, +0.022] & 0.75 & 1 & -0.048 & 0.013 \\
L2-init LoRA & Ring 3 & 10 & 0/10/0 & +0.000 & [+0.000, +0.000] & 1 & 1 & +0.000 & 0\\
Aggressive (r=16 α=32 all-proj) & Ring 1 & 21 & 2/12/7 & -0.059 & [-0.140, +0.027] & 0.203 & 0.18 & -0.238 & 0.584 \\
Aggressive (r=16 α=32 all-proj) & Ring 3 & 10 & 0/4/6 & -0.400 & [-0.650, -0.200] & 0.0312 & 0.0312 & -0.600 & 0.99 \\

Aggressive (r=32 α=64 all-proj) & Ring 1 & 21 & 2/12/7 & -0.051 & [-0.135, +0.037] & 0.359& 0.18 & -0.238 & 0.508 \\
Aggressive (r=32 α=64 all-proj) & Ring 3 & 10 & 0/8/2 & -0.150 & [-0.400, +0.000] & 0.5 & 0.5 & -0.200 & 0.813 \\

\bottomrule
\end{tabular}}
\caption{Per-(method, ring) paired analysis of per-task Recall@10 differences versus the Frozen-0.6B baseline under Paper Config A (full mix), matched on query ID. Columns report the number of paired tasks $N$, wins/ties/losses (W/T/L), mean paired difference $\Delta$, a 95\% paired-bootstrap CI from 10{,}000 resamples, the exact Wilcoxon signed-rank $p$ on non-zero paired differences, the exact sign-test $p$ on wins versus losses, Cliff's $\delta = (W-L)/N$, and the TOST $p$ for equivalence within $\varepsilon = 0.05$ Recall@10. }

\label{tab:stats-test-ring1-3}

\end{table*}

The real-task slices are small (Ring 1: \(N=21\), Ring 3: \(N=10\)), therefore, we also report per-task paired differences on Recall@10 between the fine-tuning method and baseline (Qwen 3 0.6B) for both Ring 1 and Ring 3. Table~\ref{tab:per-task_ring1} shows these differences for Ring 1 (\textsc{SkillsBench})  and Table~\ref{tab:per-task_ring3} shows these differences for  Ring 3 (\textsc{TerminalBench~2}). The adapters were trained using Track B's Config A (full mix). For example, if a value is negative in a cell, it means the fine tuning method (forgetting mitigation or aggressive) has lower recall than baseline on that task (it regressed), and if the value is positive (>0) then the method has higher Recall@10 than the baselines 0.6B (it wins). The two tables also  show {mean $\Delta$} of paired differences. The mean paired differences are 0 for Ring 3 (Table~\ref{tab:per-task_ring3})  and less than -0.005 for Ring 1 (Table~\ref{tab:per-task_ring1}). However, for aggressive methods, the mean paired differences ranged from -0.051 to -0.250. This again shows the significant regression of aggressive methods.

Table~\ref{tab:stats-test-ring1-3} summarizes the per-task Recall@10 differences. Each row pairs a fine-tuned method against frozen-0.6B along three axes: (1) \emph{Direction} is given by W/T/L (wins/ties/losses of finetuned method vs. baseline) and the mean $\Delta$ difference in Recall@10 between the fine-tune method and baseline. Positive values favour the method, negative values favour the frozen baseline. (2) \emph{Effect Size} is given by the 95\% paired-bootstrap CI on $\Delta$ (10,000 resamples) and by Cliff's $\delta = (W-L)/N$, where $|\delta|$ near $0.1$, $0.3$, and $0.5$ correspond to small, moderate, and large paired win/loss skews. (3) \emph{Strength of evidence} is given by three $p$-values that test different nulls: the exact Wilcoxon signed-rank test asks whether the non-zero paired differences are systematically shifted from zero, the exact sign test asks whether wins and losses are imbalanced beyond chance, and TOST asks whether $\Delta$ lies inside a pre-specified equivalence band of $\pm 0.05$ Recall@10.

Wilcoxon and sign tests share the standard reading in which a small $p$ is evidence of a directional change; however, TOST with a small $p$ is evidence that any residual difference is practically negligible. We claim \emph{regression} only when $\Delta < 0$ and either Wilcoxon or sign rejects at $\alpha = 0.05$, and we claim \emph{equivalence} to frozen-0.6B only when TOST rejects at $\alpha = 0.05$. All other cases are treated as ambiguous. 

Under this interpretation, the regularized methods (conservative anchor, LwF, EWC, and L2-init) are practically equivalent to frozen-0.6B on both rings, including exact ties on all Ring~3 tasks, whereas the unregularized aggressive $r=16$ all-projection model shows clear regression on Ring~3 ($0/4/6$, $\Delta=-0.400$, Wilcoxon/sign $p=0.0312$). The aggressive $r=32,\alpha=64$ model is worse on average on Ring~3 but does not meet the regression criterion ( Wilcoxon/sign $p=0.5$), so it should be interpreted as inconclusive rather than equivalent. Given the small real-task sample sizes ($N=21$ and $N=10$), ambiguous rows should be read as lack of sufficient evidence, not evidence of no effect.

\section{Out of Distribution Evaluation on Large Sample Size}
\label{appendix:large_sample_size}

Due to the small sample size of Ring 1 (N=21) and Ring 3 (N=10), Wilcoxon and the sign test showed ambiguity despite the negative delta differences for Recall@10 for Table~\ref{tab:stats-test-ring1-3}. We also evaluated aggressive and four forgetting mitigation approaches for  fine-tuning on a larger and general purpose BEIR benchmark dataset~\cite{BEIR_etal}. The BEIR benchmark is not a specialized task to skills retrieval benchmark, which this study focuses on, but due to lack of task-skills benchmarks, we use BEIR, a general purpose retrieval benchmark.   BEIR measures zero-shot retrieval performance, making it a strong indicator of real-world robustness and OOD (out-of-distribution) generalization. BEIR has 18 diverse datasets, we randomly selected 6 datasets for evaluation, namely: ArguAna (8.67K Corpus), FiQA(57K Corpus), NFCorpus (3.6K Corpus), SciDocs (25K Corpus), SciFact (5K Corpus) and Quora (0.52M Corpus). Our approach for evaluation is as follows: (a) we evaluate the Qwen3-0.6B embedding model directly on these BEIR datasets; (b) we evaluate aggressively fine-tuned LoRA  on Track B's Config A (as in Table~\ref{tab:rings}) on these datasets; (c) we evaluate LoRA fine-tuned adapter with  conservative anchor ($\lambda\!=\!0.1$),  LwF, L2-init and EWC forgetting mitigation approaches on these BEIR datasets. Finally, we present Recall@10 and MRR@10 in Table~\ref{tab:beir_results}. We can see the same pattern of recall here as in Table~\ref{tab:p7-forget-recall10}. 

\begin{table*}[htbp]
\centering
\small
\setlength{\tabcolsep}{4pt}
\begin{tabular}{llccccccc}
\toprule
Dataset & Metric & Frozen 0.6B  & Cons. anchor (0.1) & L2-init & EWC & LwF & Aggr. (r=16) & Aggr. (r=32) \\
\midrule
\multirow{2}{*}{ArguAna}
& Recall@10 & 0.9310 & 0.9346 & 0.9343 & 0.9322 & 0.9345 & 0.9047 & 0.9182 \\
& MRR@10    & 0.5880 & 0.5898 & 0.5896 & 0.5890 & 0.5864 & 0.5700 & 0.5743 \\
\midrule
\multirow{2}{*}{FiQA}
& Recall@10 & 0.4756 & 0.4698 & 0.4648 & 0.4658 & 0.4671 & 0.3651 & 0.3755 \\
& MRR@10    & 0.4682 & 0.4652 & 0.4638 & 0.4643 & 0.4674 & 0.3647 & 0.3788 \\
\midrule
\multirow{2}{*}{NFCorpus}
& Recall@10 & 0.1619 & 0.1645 & 0.1630 & 0.1633 & 0.1633 & 0.1432 & 0.1357 \\
& MRR@10    & 0.5333 & 0.5320 & 0.5316 & 0.5325 & 0.5314 & 0.5005 & 0.4576 \\
\midrule
\multirow{2}{*}{SciDocs}
& Recall@10 & 0.2047 & 0.2059 & 0.2055 & 0.2056 & 0.2055 & 0.1504 & 0.1565 \\
& MRR@10    & 0.3247 & 0.3228 & 0.3226 & 0.3220 & 0.3230 & 0.2355 & 0.2497 \\
\midrule
\multirow{2}{*}{SciFact}
& Recall@10 & 0.8182 & 0.8199 & 0.8166 & 0.8163 & 0.8166 & 0.7984 & 0.7845 \\
& MRR@10    & 0.6474 & 0.6542 & 0.6461 & 0.6500 & 0.6498 & 0.6199 & 0.6163 \\
\midrule
\multirow{2}{*}{Quora}
& Recall@10 & 0.9539 & 0.9544 & 0.9543 & 0.9542 & 0.9542 & 0.9384 & 0.9324 \\
& MRR@10    & 0.8669 & 0.8684 & 0.8681 & 0.8683 & 0.8685 & 0.8500 & 0.8493 \\
\bottomrule
\end{tabular}
\caption{Recall@10 and MRR@10 on six BEIR datasets for four regularization approaches for LoRA fine-tuning and two aggressive LoRA fine-tuning strategies. For aggressive LoRA lr=2e-5 and alpha=2r, and for all regularizer approaches lr=5e-6. }
\label{tab:beir_results}
\end{table*}

The conservative anchor, LwF, L2-init and EWC generate similar Recall@10 or slightly better than the baseline model on out-of-distribution data; whereas, aggressive fine-tuning results in poorer recall and MRR. Regardless of in-distribution or out-of-distribution, conservative anchor (and other forgetting mitigation approaches) performed better on both types of data. The forgetting mitigation approaches based LoRA performed already better on in-distribution data in Table~\ref{tab:p7-forget-recall10} and show clearly better performances on large out-of-distribution data in this section than aggressive LoRA. This provides  substantial evidence for the use of forgetting mitigation approaches in skill retrieval and other retrieval domains.

\section {Deployment}
\label{appendix:deployment}

\begin{table}[h]
\centering
\footnotesize
\setlength{\tabcolsep}{4pt}
\resizebox{\linewidth}{!}{%
\begin{tabular}{@{}l p{0.62\linewidth}@{}}
\toprule
\textbf{Metric} & \textbf{Value} \\
\midrule
End-to-end latency (p50)   & $169\,\text{ms}$ \\
End-to-end latency (p95)   & $175\,\text{ms}$ \\
End-to-end latency (mean)  & $169\,\text{ms}$ \\
Sequential QPS             & $5.91$ \\
\midrule
\multicolumn{2}{l}{\textit{Hardware}} \\
\quad Retriever  & Qwen3-Emb-0.6B + Config~A LoRA, bf16, GPU \\
\quad FAISS index & \texttt{IndexFlatIP}, 34{,}396$\times$1024 fp32, CPU (${\approx}141\,\text{MB}$) \\
\quad Reranker    & Qwen3-Rerank-0.6B + LwF LoRA ($\lambda\!=\!0.1$), bf16, GPU \\
\quad Pool size   & top-30 candidates per query \\
\midrule
Peak GPU memory (after warmup) & $4{,}296\,\text{MiB}$ / $95{,}830\,\text{MiB}$ (4.5\%) \\
\bottomrule
\end{tabular}
}
\caption{MCP server latency and resource utilisation measured on a single NVIDIA H100~SXM5 ($96\,\text{GB}$) in \texttt{gpu} mode.}
\label{tab:mcp_latency}
\end{table}

We deploy the final system as a stateless streamable HTTP Model Context
Protocol (MCP) server. The server exposes four tools:
\texttt{retrieve\_skills}, \texttt{retrieve\_skills\_debug},
\texttt{get\_skill\_detail}, and \texttt{health}, and runs the same
three-stage pipeline used in our experiments: \texttt{Qwen3-Embedding-0.6B} retriever fine-tuned with the conservative anchor
regularizer ($\lambda=0.1$), exact nearest-neighbor search with embeddings over a
\texttt{FAISS IndexFlatIP} built on \num{34396} skills, and the final
\texttt{Qwen3-Reranker-0.6B} LoRA reranker trained with listwise LwF
($\lambda=0.1$). The retriever and reranker scores are blended with the
same per-query min-max normalization and $\alpha=0.55$ used in the
deployed configuration.

On a single NVIDIA H100, the full query path (retriever encoding,
FAISS search, reranking, and score blending) completes in
$169\,\mathrm{ms}$ p50 and $175\,\mathrm{ms}$ p95, sustains
$5.9$ queries/s, and uses only $4.3\,\mathrm{GB}$ of GPU memory. The small GPU memory consumption makes it fit on commodity GPUs with roughly $8$--$12\,\mathrm{GB}$ of VRAM (e.g., RTX~3060~12GB, RTX~4060~Ti, or cloud T4/L4-class devices), although we only benchmarked latency on an H100. We also benchmarked the same MCP server in CPU-only mode on the
same VM using an AMD EPYC~9V84 processor (40 cores, observed
$2.40\,\mathrm{GHz}$, $314\,\mathrm{GiB}$ RAM). On a 100-query
sequential trace from the held-out synthetic evaluation set, CPU mode
achieves $8.48\,\mathrm{s}$ p50 and $12.89\,\mathrm{s}$ p95 latency,
with mean latency $9.07\,\mathrm{s}$, maximum latency
$17.03\,\mathrm{s}$, and sequential throughput
$0.11\,\mathrm{QPS}$. The loaded CPU mode server uses
approximately $6.1\,\mathrm{GiB}$ resident memory, making CPU mode
practical for correctness testing and low-throughput deployment, but
not for interactive use. This suggests that deployment is feasible on substantially cheaper hardware, trading off throughput and response time. This is also summarized in Table~\ref{tab:mcp_latency}.

\end{document}